\documentclass[rnote]{aa} 

\usepackage{graphicx}
\usepackage{longtable}
\usepackage{natbib}
\usepackage{amsmath}
\usepackage{amsfonts}
\usepackage{amssymb}
\usepackage{url}

\bibpunct{(}{)}{;}{a}{}{,}

\begin{document}

\title{Preliminary determinations of the masses of the neutron star and mass donor in the High Mass X-Ray Binary system  {EXO~1722--363} \thanks{Based on observations carried out at the European Southern Observatory under programme ID 077.B-0872(A)}}

\author{A.B. Mason \inst{1}
\and A.J. Norton   \inst{1}
\and J.S. Clark  \inst{1}
\and I. Negueruela \inst{2}
\and P. Roche \inst{1,3,4}}

\institute{Department of Physics \& Astronomy, The Open University, Milton Keynes MK7 6AA, UK \and
 Departamento de F\'{\i}sica, Ingenier\'{\i}a de Sistemas y
  Teor\'{\i}a de la Se\~{n}al, Universidad de Alicante, Apdo. 99,
  E03080 Alicante, Spain \and
  School of Physics \& Astronomy, Cardiff University, The Parade, Cardiff, CF24 3AA, UK. \and
  Division of Earth, Space \& Environment, University of Glamorgan, Pontypridd, CF37 1DL, UK.}

\date{Received 01 October 2009 / Accepted 17 November 2009}

\abstract{}
{We intended to measure the radial velocity curve of the supergiant companion to the eclipsing high mass X-ray binary pulsar  {EXO~1722--363} and hence determine the stellar masses of the components.}
{We used a set of archival K$_{\rm s}$-band infrared spectra of the counterpart to  {EXO~1722--363} obtained using ISAAC on the VLT, and cross-correlated them in order to measure the radial velocity of the star.}
{The resulting radial velocity curve has a semi-amplitude of $24.5 \pm 5.0$~km~s$^{-1}$. When combined with other measured parameters of the system, this yields masses in the range 1.5 $\pm$ 0.4 - 1.6 $\pm$ 0.4 ~M$_{\odot}$ for the neutron star and 13.6 $\pm$ 1.6 - 15.2 $\pm$  1.9 ~M$_{\odot}$ for the B0--1 Ia supergiant companion. These lower and upper limits were obtained under the assumption that the system is viewed edge-on (i = 90$^\circ$) for the lower limit and the supergiant fills its Roche lobe ($\beta = 1$) for the upper limit respectively. The system inclination is constrained to $i>75^{\circ}$ and the Roche lobe-filling factor of the supergiant is $\beta>0.9$.
Additionally we were able to further constrain our distance determination to be 7.1 $\le$ d $\le$ 7.9 kpc for  {EXO~1722--363}. The X-ray luminosity for this distance range is 4.7 $\times$ 10$^{35}$ $\le$ L$_{\rm X}$ $\le$ 9.2 $\times$ 10$^{36}$ erg s$^{-1}$.} 
{ {EXO~1722--363} therefore becomes the seventh of the ten known eclipsing X-ray binary pulsars for which a dynamical neutron star mass solution has been determined. Additionally  {EXO~1722--363} is the first such system to have a neutron star mass measurement made utilising near-infrared spectroscopy.}
{}


\authorrunning{A.B. Mason et al}
\titlerunning{Preliminary mass of the NS and donor in HMXB EXO1722-363} 

\maketitle

\section{Introduction}

The precise form of the neutron star (NS) equation of state is still unknown. Despite much theoretical work  aimed at determining this fundamental aspect of astrophysics, to eliminate some of the contending theories we must turn to observational data. Presently the only means of determining the mass of neutron stars in accretion driven systems is by observing eclipsing X-ray binary pulsars. Unfortunately only 10 such systems are currently known, and only 6 of these have previously had mass measurements made (e.g. \cite{ash99}, \cite{quaintrell03}, \cite{valbaker05}).  In this paper we present the preliminary results from our on-going work on the mass of the High Mass X-ray Binary (HMXB) accretion driven pulsar  {EXO 1772--363}. The counterpart star within this HMXB is heavily obscured and reddened, necessitating for the first time the utilisation of near-infrared spectroscopy to construct the Radial Velocity (RV) curve in order to obtain an accurate mass solution.

Observations made of  {EXO~1722--363} (alternatively designated IGR J17252--3616), in 1987 by the \emph{Ginga} X-ray satellite were the first to detect pulsations. These pulsations were found to have a $413.9 \pm 0.2$s period \citep{tawara89}. Over an 8 hour period the source appeared to vary substantially in flux, decreasing from 2~mcrab to 0.2 -- 0.3~mcrab within the 6--21~keV band. It was subsequently found that the X-ray flux remained persistent from 20 -- 60~keV, however a cutoff point was found above this flux level in which the source was undetectable.  {EXO~1722--363} at maximum flux and assuming a distance of 10 kpc, had a luminosity calculated as $5 \times 10^{36}$~erg~s$^{-1}$ \citep{tawara89}.
Later observations by the \emph{Rossi X-ray Timing Explorer} (RXTE) revealed the eclipsing nature of this system with the eclipse duration determined as $1.7 \pm 0.1$~days \citep{corbet05}. Subsequent observations by \emph{INTEGRAL} followed up by \emph{XMM-Newton} in 2004 led to a further refinement of the spin and orbital periods to $413.851\pm0.004$~s and $9.7403\pm0.0004$~days respectively \citep{thompson07}.

\emph{XMM-Newton} observations allowed the source position to be determined more precisely (with an uncertainty of $4^{\prime\prime}$) at RA(2000.0) = $17^h 25^m 11.4^s$ and Dec = $-36^\circ 16 ^\prime 58.6 ^{\prime\prime}$.  {EXO~1722--363} lies within the Galactic plane and as it is heavily reddened, unsurprisingly the counterpart star could not be detected optically. An infrared counterpart was found lying $1^{\prime\prime}$ from the X-ray source position \citep{zurita06} with a corresponding entry in the 2MASS catalogue, 2MASS~J17251139--3616575 (with JHK magnitudes J = 14.2, H = 11.8 and K$_s$ = 10.7). Examination of near infrared K-band spectra obtained with the ESO \emph{ISAAC} instrument led to our determination of the spectral classification of the mass donor as B0 -- B1 Ia \citep{mason09}.

\section{Observations and Data Reduction}

In our previous work, we only had single epoch K$_s$-band spectra of the mass donor in  {EXO~1722--363}, but in order to determine a dynamical mass solution, radial velocities at a range of orbital phases are required. Fortunately we were able to locate a series of K$_{s}$-band spectra held within the ESO Archive\footnote{\url {http://archive.eso.org/eso/eso\_archive_main.html}} which were obtained over 26 nights between 24th May, 2006 and 4th August, 2006. 26 pairs of spectra were centred on 2.1~$\mu$m and 26 pairs centred on 2.2~$\mu$m. No observations of radial velocity standards appear to have been taken with any of the science spectra, however there are telluric standards available to enable removal of atmospheric features from the target spectra.

The data were reduced using the ISAAC pipeline \footnote{\url {http://www.eso.org/sci/data-processing/software/pipelines/isaac/isaac-pipe-recipes.html}} in conjunction with the data browsing tool GASGANO \footnote{\url {http://www.eso.org/sci/data-processing/software/gasgano/}}. Unfortunately within the~2.2 $\mu$m  {EXO 1722-363} dataset, the counterpart to  {EXO~1722--363} had been incorrectly identified and the telescope had been mis-pointed, so the spectra were unusable. Additionally only 11 of the 26 spectra from the dataset centred on 2.1~$\mu$m turned out to be of sufficient quality to derive a radial velocity measurement (See Table~\ref{usable_spectra}).
The usable spectra were made in the SW MRes mode with a 0.6$^{\prime\prime}$ slit. Resulting in spectra with a high S/N and resolution (R $\approx$ 4200). The integration time for each pair of spectra was 700~s, with the resulting data having a count rate below 10\,000 ADU; therefore no correction for non-linearity was necessary.

\begin{figure}
    \includegraphics[width=9cm]{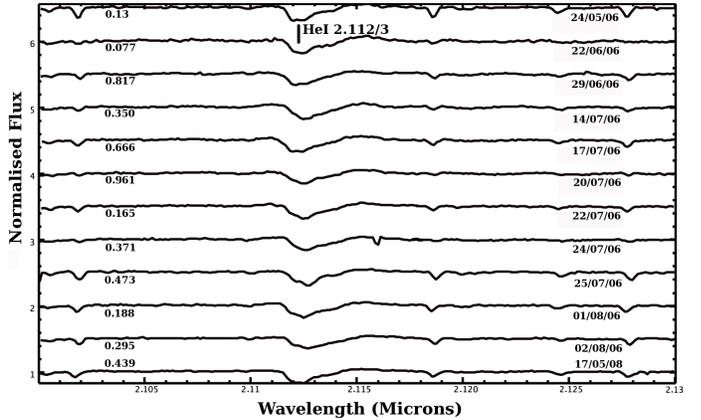}
    \begin{center}
    \caption{Continuum normalised K$_{s}$-band spectra centred on 2.1~$\mu$m of  {EXO~1722--363} in order of date from top to bottom.}
    \label{stacked_spectra}
    \end{center}
\end{figure}

Flatfields were reduced using the pipeline recipe isaac\_spc\_flat and combined to produce a master flatfield. The wavelength calibration and ISAAC slit curvature distortion was computed using OH skylines using the pipeline recipe isaac\_spc\_arc. Spectra produced by ISAAC have a high degree of curvature, to remove this the pipeline recipe isaac\_spc\_startrace computes the spectra curvature using both images and spectra of a star moving across the slit. Science spectra were obtained using the nodding technique; unfortunately only one nod was performed during the original observation, which we believe to be less than optimal. The two nodded science frames were then reduced using the products of the pipeline calibration recipes to produce a final reduced science spectrum. This process was then repeated for each telluric standard. Telluric correction was then made using the standards shown in Table~\ref{usable_spectra}.
All spectra were reduced using standard IRAF\footnote{IRAF is distributed by the National Optical Astronomy Observatory, which is operated by the Association of Universities for Research in Astronomy, Inc., under cooperative agreement with the National Science Foundation.} routines; Fig~\ref{stacked_spectra} shows the stacked continuum normalised spectra ordered by date of observation.

\section{Data Analysis}

Radial velocities were determined by cross-correlating the region around the HeI 2.112/3~$\mu$m absorption line in each of the 11 archive spectra against the high signal-to-noise K$_s$-band spectrum of EXO 1722-363, which we had previously obtained for spectral classification purposes \citep{mason09}. The resulting velocities were then  corrected to the solar system barycentre and are reported in Table~\ref{usable_spectra}. The spectrum highlighted in bold is that used as the reference spectrum (our previously obtained high signal-to-noise K$_s$-band spectrum of EXO 1722-363 \citep{mason09}). In obtaining these final velocities, we determined the absolute velocity of the reference spectrum by fitting the positions of its absorption lines.

\begin{table*}
\caption{The phase, radial velocity and telluric standard for each  {EXO~1722--363} archive spectrum.}
\label{usable_spectra}
\centering
\begin{tabular} {cccccc}
\hline
Mid-point of Observations (UT) & HJD & Phase & Radial velocity / km s$^{-1}$ & Telluric Std & Telluric Spec. Type \\
\hline
 2006 May 24.340 & 245 3879.83 & 0.130 & --18.61 $\pm$ 11.8  & Hip 093225 & B4V \\
 2006 Jun 22.025 & 245 3908.53 & 0.077 & --11.74 $\pm$ 11.8  & Hip 088109 & B5II\\
 2006 Jun 29.233 & 245 3915.74 & 0.817 & --21.39 $\pm$ 11.8 & Hip 070148 & B8III\\
 2006 Jul 14.167 & 245 3930.67 & 0.350 & 14.92 $\pm$ 11.8 & Hip 089960 & B6V\\
 2006 Jul 17.248 & 245 3933.75 & 0.666 & --29.22 $\pm$ 11.8 & Hip 087616 & B9IV/V\\
 2006 Jul 20.122 & 245 3936.63 & 0.961 & --14.95  $\pm$ 11.8 & Hip 090336 & B7III\\
 2006 Jul 22.112 & 245 3938.62 & 0.165 & --10.09  $\pm$ 11.8 & Hip 094859 & B5V\\
 2006 Jul 24.115 & 245 3940.62 & 0.371 & 21.21  $\pm$ 11.8 & Hip 090336 & B7III\\
 2006 Jul 25.105 & 245 3941.61 & 0.473 & 6.89 $\pm$ 11.8 & Hip 089960 & B6V\\
 2006 Aug 01.111 & 245 3948.57 & 0.188 & --3.83 $\pm$ 11.8 & Hip 085548 & B9II\\
 2006 Aug 02.113 & 245 3949.62 & 0.295 & 28.06 $\pm$ 11.8 & Hip 085548 & B9II\\
\bf {2008 May 17.115} & \bf {245 4603.62} & \bf{0.439} & \bf{30.95 $\pm$ 11.8} & \bf{Hip 085008} & \bf{B5V} \\
\end{tabular}
\end{table*}

From X-ray data there is no evidence that  {EXO~1722--363} has anything other than a circular orbit \citep{thompson07}, so we fitted the radial velocities of the supergiant star with a simple sinusoidal solution. The ephemeris of \citet{thompson07} specifies the epoch of mid-eclipse as
\begin{equation}
   T({\rm HJD}) = 53761.68(4) + 9.7403(4)N
\end{equation}
where $N$ is the cycle number and uncertainties in brackets refer to the last decimal place quoted. At the epoch of our observations, the accumulated uncertainty in phase is formally only $\sim 0.005$, but nonetheless we fitted our data with two models, in one of which the zero phase was a free parameter and in the other of which it was not.

\begin{figure}[h] 
 \includegraphics[scale=0.33, angle=-90]{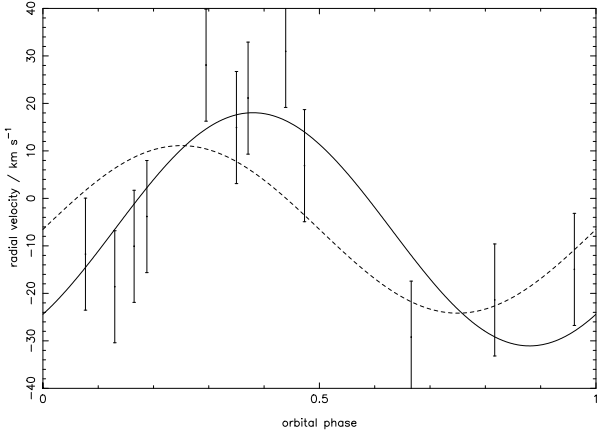}
    \begin{center}
    \caption{Radial velocity data for the supergiant star in  {EXO~1722--363}. The solid line is the best fitting sinusoid with three free parameters, the dashed line is that with a fixed zero phase in line with the published ephemeris. The orbital phase is based upon the ephemeris of \citet{thompson07}.}
    \label{rvcurve}
    \end{center}
\end{figure}

Fitting our data with a sinusoid with just two free parameters (RV amplitude and systemic velocity) yielded an amplitude of 17.6 $\pm$ 7.7~km~s$^{-1}$ and a systemic velocity of  -6.5 $\pm$ 5.6~km~s$^{-1}$. In order to achieve a reduced chi-squared of unity, the uncertainties on each RV data point had to be scaled to $\pm$ 17.5~km~s$^{-1}$. In comparison, fitting our data with a sinusoid with three free parameters (i.e. with the addition of zero phase as a free parameter) gave an amplitude of 24.6 $\pm$ 5.0~km~s$^{-1}$, a systemic velocity of $-6.5 \pm 3.8$~km~s$^{-1}$ and a phase shift of -0.13 $\pm$ 0.03. In this case, a reduced chi-squared of unity was achieved by scaling the uncertainties on each point to $\pm$ 11.8~km~s$^{-1}$.  Although the best-fit phase offset is discrepant with the accumulated phase uncertainty of the ephemeris, we prefer this fit and use the data from it subsequently; both fits are shown in Figure~\ref{rvcurve} in which the value for $K_{\rm O}$ is that resulting from fitting the radial velocities including a phase shift.

The masses of the system components may be determined as follows. The mass ratio of the system $q$ is equal to the ratio of the semi-amplitudes of the radial velocities for each star
\begin{equation}
     q = \frac{M_{\rm X}}{M_{\rm O}} = \frac{K_{\rm O}}{K_{\rm X}}
\end{equation}
where $M_{\rm X}$ and $M_{\rm O}$ are the masses of the neutron star and supergiant star respectively, and
$K_{\rm X}$ and $K_{\rm O}$ are the corresponding semi-amplitudes of their radial velocities. In addition, for circular orbits,
\begin{equation}
M_{\rm O} = \frac{{K_{\rm X}}^3 P}{2\pi G \sin^3 i}\left(1+q\right)^2
\end{equation}
and similarly
\begin{equation}
M_{\rm X} = \frac{{K_{\rm O}}^3 P}{2 \pi G \sin^3 i}\left(1+\frac{1}{q} \right)^2
\end{equation}
where $i$ is the inclination to the plane of the sky and $P$ is the orbital period. For  {EXO~1722--363}, X-ray pulse timing delays yield the value of $K_{\rm X}$ as $226.1\pm6.7$~km~s$^{-1}$ \citep{thompson07}. A value for the system inclination can be found from the geometric relation
\begin{equation}
     \sin i \approx \frac{\left[1 - \beta^2 \left(\frac{R_L}{a} \right)^2\right]^{1/2}}{\cos~\theta_{\rm e}}
\end{equation}
where $\theta_{\rm e}$ is the eclipse half-angle, $R_{\rm L}$ is the Roche lobe radius of the supergiant, $\beta$ is the ratio the supergiant's radius to that of its Roche lobe and $a$ is the separation between the centres of mass of the two stars. The Roche lobe radius may be approximated by
\begin{equation}
   \frac{R_{\rm L}}{a} \approx A + B \log q + C \log^2 q
\end{equation}
where the constants have been determined by \citet{rappaport84} as
\begin{equation}
A \approx 0.398 - 0.026\Omega^2 + 0.004\Omega^3
\end{equation}
\begin{equation}
B \approx - 0.264 + 0.052\Omega^2 - 0.015\Omega^3
\end{equation}
\begin{equation}
C \approx - 0.023 - 0.005\Omega^2
\end{equation}
$\Omega$ is the ratio of the spin period of the supergiant to its orbital period. For  {EXO~1722--363} we have assumed that the supergiant is close to Roche lobe-filling and is rotating synchronously with the orbit, so $\Omega =1$ (although we note this assumption may not be entirely correct in this case), and the eclipse half angle is measured using RXTE data to be $\theta_{\rm e} = 31.8^{\circ} \pm 1.8^{\circ}$ \citep{corbet05}.

Hence the above set of equations allow the masses of the two stars to be determined in two limits. First, assuming that the supergiant fills its Roche lobe (in which case $\beta=1$) we can find a lower limit to the system inclination $i$ and upper limits to the stellar masses. Secondly, assuming that the system is viewed edge-on (in which case $i=90^{\circ}$) we can find a lower limit to the Roche lobe filling factor $\beta$ and lower limits to the stellar masses. Unfortunately, the spectra we have obtained from the ESO archive are not of sufficient quality to conduct a non-LTE model atmosphere analysis and hence make an accurate determination of the stellar radius which would break this degeneracy.

In order to propagate the uncertainties in each parameter, we performed a Monte Carlo analysis of the above equations to determine the system masses. The results in each limit are shown in Table~\ref{results} , and of course masses lying between the extremes are also valid and correspond to values of $i$ and $\beta$ between their extremes.

\begin{table}
\caption{System parameters for  {EXO~1722--363}.} 
 \label{results}
 \begin{tabular}{llll} \hline
Parameter & \multicolumn{2}{c}{Value} & Ref. \\ \hline
{\it Observed}     		&	&	& \\
$a_{\rm X} \sin i$ / lt sec	& \multicolumn{2}{c}{$101 \pm 1$}		& [1]\\
$P$ / d				& \multicolumn{2}{c}{$9.7403 \pm 0.0004$}	& [1]\\
$T_{90}$ / HJD		& \multicolumn{2}{c}{$53761.68 \pm 0.04$}	        & [1]\\	
$e$				& \multicolumn{2}{c}{$<0.19$}	& [1]\\	
$\theta_{\rm e}$ / deg		& \multicolumn{2}{c}{$31.8 \pm 1.8$}		& [2]\\	

$K_{\rm O}$ / km s$^{-1}$	& \multicolumn{2}{c}{$24.5 \pm 5.0$} 		& [3]\\

{\it Assumed}     		&	&	& \\
\bf{$\Omega$}			& \multicolumn{2}{c}{= 1}		     \\

{\it Inferred} & & & \\

$K_{\rm X}$ / km s$^{-1}$	& \multicolumn{2}{c}{$226.1 \pm 6.7$}  		\\
$q$				& \multicolumn{2}{c}{$0.107 \pm 0.022$}  \\
$\beta$				& $1.000$ 	& $0.916 \pm 0.047$	 \\
$i$ / deg			& 75.2 $\pm ~4.6$& $90.0$                 \\
$M_{\rm X}$ / M$_{\odot}$ 	& 1.63 $\pm ~0.38$ & $1.46 ~\pm$ 0.38 	 \\
$M_{\rm O}$ / M$_{\odot}$ 	& 15.2 $\pm $ 1.9 & $13.6 \pm 1.6$ 	 \\
$a$ / R$_{\odot}$ & $49.1 ~\pm$ 9.1 & $47.3 \pm 8.8$  \\
$R_{\rm L}$ / R$_{\odot}$	& 28.0 $\pm$ 5.3  & $27.0 ~\pm$ 5.0   \\
$R_{\rm O}$ / R$_{\odot}$	& 28.0 $\pm$ 5.3  & $24.7 ~\pm$ 4.7           \\ \hline
\end{tabular}\\
$[1]$ Thompson et al. 2007; $[2]$ Corbet et al. 2005\\
$[3]$ this paper
\end{table}

\section{Discussion}

\begin{figure}
    \includegraphics[width=9cm]{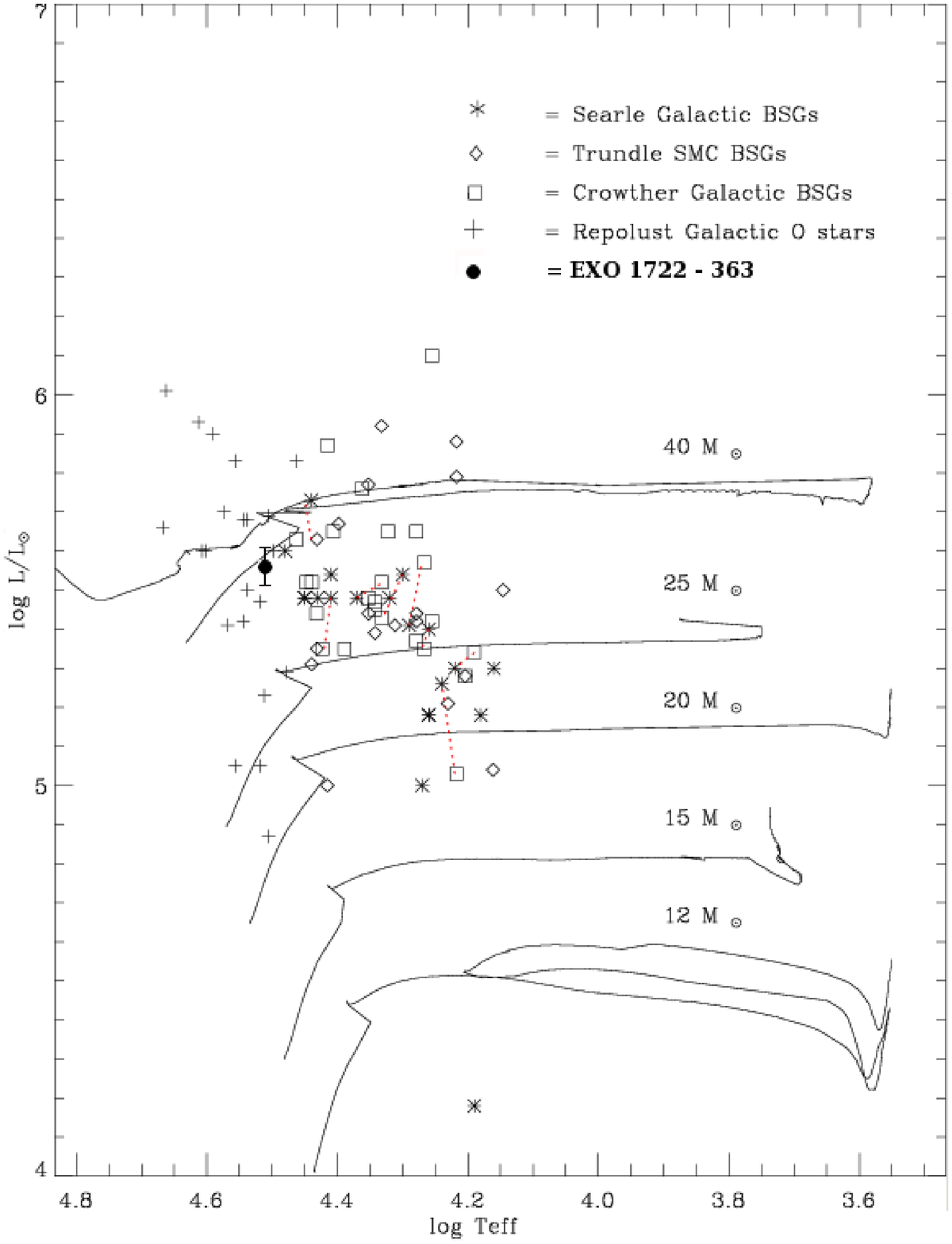}
    \begin{center}
    	\caption{Position of  {EXO 1722-363} on the Hertzsprung-Russell \citep{searle08} diagram alongside a sample of O and B supergiants 	from differing locations, Galactic B supergiants, \cite{crowther06}, SMC B supergiants (\cite{trundle04};          \cite{trundle05}) and Galactic O stars, \cite{repolust04}. These are overplotted together with solar metallicity evolutionary tracks from \cite{meynet00}. Also shown is the lower and upper limits on the luminosity of  {EXO 1772-363}.} 
        \label{evo_track}
    \end{center}
\end{figure}

Although the 11 spectra reported here are of relatively low quality, and few in number, they still allow us to make a
 preliminary determination of the orbit of the supergiant in  {EXO~1722--363} and make a first measurement of the 
dynamical masses of the stellar components. The results are encouraging for a number of reasons. First, the resulting
 neutron star mass is consistent with the canonical mass of 1.4~M$_{\odot}$ measured in most other eclipsing HMXBs, 
except for that in Vela X-1, \citep{quaintrell03}. Second, the measured mass and radius of the supergiant, $M \sim 
13 - 15$~M$_{\odot}$ and $R \sim 25 - 28$~R$_{\odot}$, support the B0-1 Ia spectral classification that we have 
previously determined \citep{mason09}. This is illustrated by the Hertzsprung-Russell diagram plotted in 
Fig. \ref{evo_track}, which shows a close correspondence between the system primary and the properties of other galactic 
field BSGs \citep{searle08}. While the similarity in temperature is to be expected - the value for the primary was 
adopted on the basis of its spectral type, which in turn has been calibrated by the analysis of \cite{searle08}, 
the radii for the field BSGs were determined via non-LTE model atmosphere analysis, while that for the primary is 
instead determined dynamically.

The measurement of the stellar radius and hence bolometric luminosity,
  has in turn has allowed a more precise determination of the distance 
to the system by comparison to its observed photometric magnitude and reddening. The refined distance to 
 {EXO~1722--363} of 7.1 - 7.9 kpc results in  an X-ray luminosity ranging from 
{L$_{\rm {X_{min}}}$ = 0.47} 
$\times$ ~10$^{36}$ to 
L$_{\rm {X_{max}}}$ = 9.2 
$\times$ ~10$^{36}$ erg s$^{-1}$. 
However,  due to the nature of the archive observations
 used for this work, the uncertainties on the mass and radius parameters are still rather large;  it 
is our intention in the  near future to propose and obtain more accurate VLT/ISAAC observations to further
 constrain the orbital solution parameters for this HMXB system.   

Finally, comparison to evolutionary tracks in Fig. \ref{evo_track} might suggest the primary had an initial progenitor
mass of $\sim$ 35 - 40M$_{\odot}$ and hence the neutron star originated in a more massive star. However we caution
that the binary is highly likely to have undergone at least one episode of mass transfer in the past, rendering such a 
conclusion highly uncertain. As an exemplar we cite  {GX 301-2}, a HMXB composed of a NS and a B hypergiant with a 
spectroscopic mass of 43$\pm$10M$_{\odot}$ \citep{kaper06}. 
 However \citep{wellstein99} propose a formation scenario in which two stars of comparable initial masses evolved via quasi conservative mass transfer into the current configuration  post supernova; hence determining progenitor masses for both primary and neutron star based on the current system parameters is non trivial.

\begin{acknowledgements}
ABM acknowledges support from an STFC studentship. JSC acknowledges support from an RCUK fellowship.
This research is partially supported by grants AYA2008-06166-C03-03 and
Consolider-GTC CSD-2006-00070 from the Spanish Ministerio de Ciencia e
Innovaci\'on (MICINN).
Based on observations carried out at the European Southern Observatory, Chile through programme ID 077.B-0872(A).
\end{acknowledgements}

\bibliographystyle{aa}
\bibliography{first_paper}

\end{document}